\documentclass[a4paper,11pt]{article}
\usepackage{pos}
\usepackage{arydshln}
\usepackage{comment}
\usepackage{subcaption}

\title{Exploring the Discovery Reach for a 95 GeV Scalar in Future $e^+e^-$ Collisions}

\author*[a]{Mukesh Kumar}
\author[a,b]{Pramod Sharma}
\author[a]{Karabo Mosala}
\author[a,c]{Bruce Mellado}

\affiliation[a]{School of Physics and Institute for Collider Particle Physics, University of the Witwatersrand, Johannesburg, Wits 2050, South Africa.}

\affiliation[b]{Indian Institute of Science Education and Research, Knowledge City, Sector 81, S. A. S. Nagar, Manauli PO 140306, Punjab, India.}

\affiliation[c]{iThemba LABS, National Research Foundation, PO Box 722, Somerset West 7129, South Africa.}

\emailAdd{mukesh.kumar@cern.ch}
\emailAdd{pramodsharma.iiser@gmail.com}
\emailAdd{bmellado@mail.cern.ch}

\abstract{The observed indications for a new scalar resonance with a mass around 95\,GeV, initially reported by LEP and supported by CMS and ATLAS in di-photon, $\tau \tau$, and $W^+ W^-$ channels, motivate exploring its discovery potential at future electron-positron colliders. This study focuses on the production of the new scalar ($S$) via $e^+ e^- \rightarrow ZS $ with $Z  \rightarrow \mu^+ \mu^- $ and $S \rightarrow b \bar{b}$ and optimizes the signal recognition using the recoil-mass method. By employing deep neural networks for signal-background discrimination, we demonstrate that a 95\,GeV scalar, mixing with the Standard Model Higgs by an angle of $\sim$0.1, can be observed with a 5$\sigma$ significance at $\sqrt{s}$ = 250\,GeV or 200\,GeV with 5~ab$^{-1}$ of integrated luminosity. 
}

\FullConference{XXXII International Workshop on Deep Inelastic Scattering and Related Subjects (DIS2025)\\
24-28 March, 2025\\
Cape Town, South Africa\\}

\begin{document}
\maketitle

\section{Introduction}
\label{intro}
The scalar particle with a mass of 125\,GeV observed by the CMS~\cite{CMS:2012qbp} and ATLAS~\cite{ATLAS:2012yve} experiments at the Large Hadron Collider (LHC) exhibits properties consistent with the ones of the predicted by the Standard Model (SM). While this discovery represents a major triumph for the SM, the underlying symmetry principles do not preclude the existence of additional Higgs bosons. Intriguingly, the CMS~\cite{CMS:2018cyk} and ATLAS~\cite{ATLAS:2024bjr} experiments have reported local excesses with significances of $2.9\sigma$ and $1.7\sigma$, respectively, near 95\,GeV in the diphoton final state. Furthermore, a global significance of $3.4\sigma$ is obtained within a combined analysis including the $\tau\tau$~\cite{CMS:2022goy} and the $WW$ channels~\cite{Coloretti:2023wng, Bhattacharya:2023lmu}.

Motivated by these hints for a new scalar boson, we investigate the discovery potential of future $e^+e^-$ colliders -- including the Circular Electron–Positron Collider (CEPC), the International Linear Collider (ILC), the Compact Linear Collider (CLIC), and the Future Circular Collider (FCC-ee). The signal process considered is $e^+ e^- \rightarrow Z^* \rightarrow Z S$, with subsequent decays $Z \rightarrow \mu^+ \mu^-$ and $S \rightarrow b \bar{b}$. The dominant background arises from the Standard Model process $e^+ e^- \rightarrow Z b \bar{b}$, with $Z \rightarrow \mu^+ \mu^-$. To enhance the signal over background sensitivity, we use the recoil mass method and a Deep Neural Network (DNN) -- based classifier.

\section{Signal and background analysis}
The dominant production mechanism for the scalar boson 
$S$ in the center-of-mass energy range of 220\,GeV to 350\,GeV is the Higgs-strahlung process, $e^+ e^- \rightarrow Z^* S$. {The $SZZ$ coupling is parametrized by a factor $\kappa_Z$ within the $\kappa$-framework and} can be probed using the recoil mass of the $Z$ boson, with the decay channel $Z \rightarrow \mu^+ \mu^-$ offering excellent sensitivity due to the precise reconstruction of muon four-momenta. The recoil mass,
\begin{align}
\label{eq:3.2}
   M_{\rm recoil} = \sqrt{s + M_{\mu^{+} \mu^{-}}^2 - 2(E_{\mu^{+}} + {E_{\mu^{-}}})\sqrt{s}}\,,
\end{align}
where $M_{\mu^+ \mu^-}$ denotes the invariant mass of the muon pair and  $E_{\mu^{\pm}}$ represent the energies of the individual muons. 

The signal and background events are generated using \texttt{MadGraph} at center-of-mass energies of $\sqrt{s} = 200\,\mathrm{GeV}$ and $250\,\mathrm{GeV}$ with the following generation-level cuts: transverse momentum of muons $p_{T_\mu} > 10\,\mathrm{GeV}$, pseudo-rapidity of muons $|\eta_\mu| < 2.5$, and angular distance between muons $\Delta R_{\mu^+ \mu^-} > 0.4$. The parton-level events are then processed through \texttt{Pythia8} for showering and hadronization, followed by \texttt{Delphes} to simulate detector effects.

\begin{figure}[htp]
\centering
\includegraphics[width=0.5\textwidth]{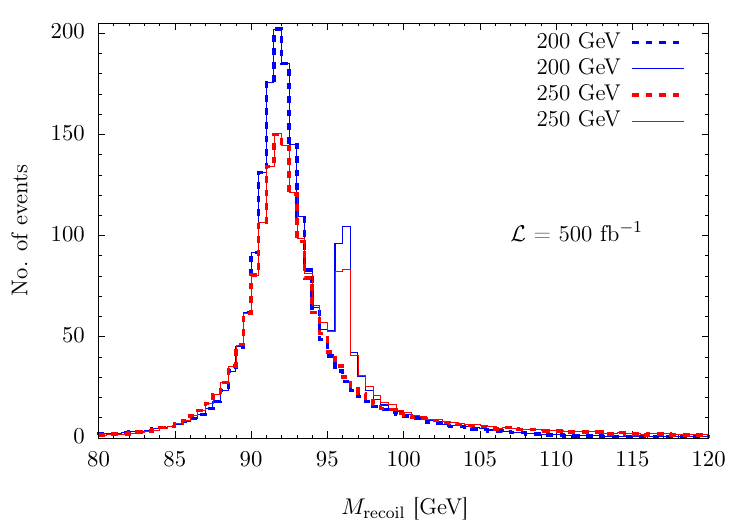}
\caption{ Recoil-mass distributions for the SM background (dashed) and the combined signal plus background (solid) for a scalar mass of $m_{S} = 95.5\,\mathrm{GeV}$ at $\sqrt{s} = 200\,\mathrm{GeV}$ (blue) and $\sqrt{s} = 250~\mathrm{GeV}$ (red) with an integrated luminosity of $\mathcal{L} = 500\,\mathrm{fb}^{-1}$.}
\label{fig:Rec}
\end{figure}

For the optimization of signal over background, we require at least two $b$-tagged jets and two muons with energies $E_{b,\mu} >$5\,GeV. The recoil mass distribution $M_{\rm recoil}$ is computed and shown in~\autoref{fig:Rec}. We then apply selection cuts on the recoil mass $M_{\rm recoil} < 120\,\mathrm{GeV}$ and on the invariant mass of the $b$-jet pair $M_{b \bar{b}} < 100\,\mathrm{GeV}$ for both signal and background events. As presented in~\autoref{tab:cutflow}, the signal significance improves from $2.4\sigma$ to $3.4\sigma$ at an integrated luminosity of $\mathcal{L} = 1~\mathrm{ab}^{-1}$ for $\sqrt{s} = 250\,\mathrm{GeV}$ and $\kappa_Z^2 = 0.1$.

\begin{table*}[htp]
    \centering
    \begin{tabular}{c|c|c|c}
    \hline
       Kinematic cuts  & $\mathcal{N}_S$  & $\mathcal{N}_B$ &  significance \\
       \hline
       \hline
    At least 2 $b$-tagged jets and 2 muons &  188 (219)  & 2926 (1876) & 2.4$\sigma$ (3.8$\sigma$)\\
    $E_{b,\mu} > $ 5\,GeV  & 188 (219)  & 2924 (1874) & 2.4$\sigma$ (3.8$\sigma$) \\
    $M_{b\bar{b}} < $ 100 GeV &   187 (217) & 2223 (1864) & 2.9$\sigma$ (3.8$\sigma$) \\
    $M_{\text{recoil}} < $ 120 GeV & 184 (216) & 1732 (1851) & 3.4$\sigma$ (3.8$\sigma$) \\ \hdashline
   93.5\,GeV $ < M_{\text{recoil}} <  $ 97.5\,GeV (before DNN) & 150 (193)  &  288 (274) & 8.4$\sigma$ (11.1$\sigma$)\\ 
   93.5\,GeV $ < M_{\text{recoil}} <  $ 97.5\,GeV (after DNN) &  54 (71) &  14 (14) &  14.5$\sigma$ (18.9$\sigma$) \\
    \hline
    \end{tabular}
    \caption{Cut-flow table showing the number of signal ($\mathcal{N}_S$) and background ($\mathcal{N}_B$) events, along with the signal significance 
    (as defined in \autoref{sig_sys}),
    after each kinematic cut for $\sqrt{s} = 250\,\mathrm{GeV}$ (200\,GeV) at $\mathcal{L} = 500\,\mathrm{fb}^{-1}$. The signal and background cross-sections are 1.1\,fb (1.4\,fb) and 16.8\,fb (12.79\,fb), respectively. The signal cross-section for the 95.5~GeV scalar is obtained by rescaling the SM Higgs cross-section by $\kappa_Z^2 = 0.1$.}
    \label{tab:cutflow}
\end{table*}

To improve the signal over background ratio, we employ a deep neural network for classification. A total of 13 kinematic observables are used as input features for training, including the energies, polar angles, and azimuthal angles of muons and $b$-tagged jets, as well as the invariant mass of the muon pair. For a detailed discussion on the DNN configuration and state-of-the-art techniques, we refer the reader to Ref.~\cite{Sharma:2024vhv}. In this study, the Area Under the Curve (AUC) scores for the training and testing samples are 97.4\% (95.2\%) and 97.2\% (95\%) for $\sqrt{s} = 200 \,\mathrm{GeV}$ and $250\,\mathrm{GeV}$, respectively. These results demonstrate that the model generalizes well from the training set to the test data.

\section{Results}
\label{results}
The signal significance is calculated by
\begin{equation}
S (\delta_{\rm sys}) = \frac{{\cal N}_\text{S}}{ \sqrt{{\cal N}_\text{B} + (\delta_{\rm sys} \cdot {\cal N}_\text{B})^{2}}}, \label{sig_sys}
\end{equation}
where $\mathcal{N}_S$ and $\mathcal{N}_B$ denote the number of signal and background events at an integrated luminosity $\mathcal{L}$, and $\delta_{\rm sys}$ represents the systematic uncertainty in the measurement. In this analysis, we consider $\delta_{\rm sys} = 2\%$.

\begin{figure*}[t]
\centering
\subfloat[\label{fig:significance}]{\includegraphics[width=0.45\textwidth,height=0.43\textwidth]{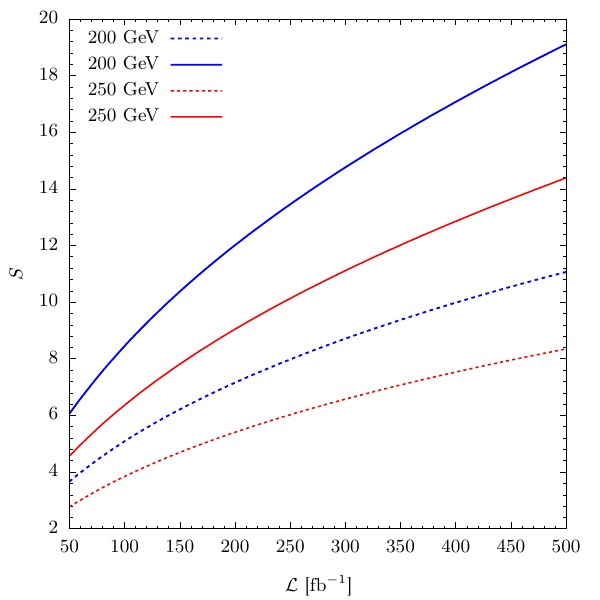}} \qquad
\subfloat[\label{fig:discovery}]{\includegraphics[width=0.45\textwidth,height=0.44\textwidth]{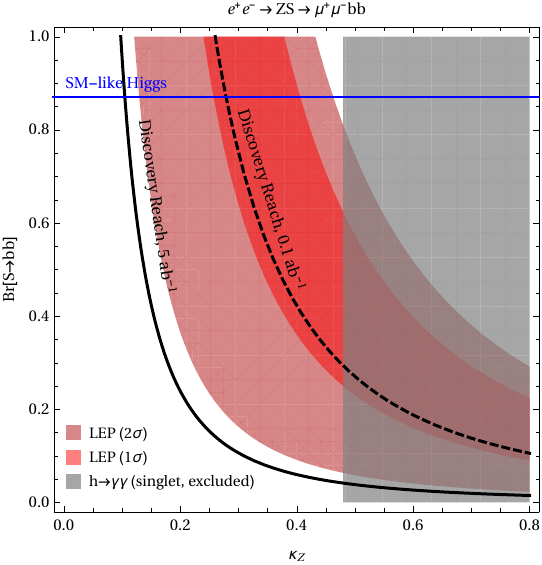}}
\caption{(a) Signal significance as a function of integrated luminosity for $m_{S} = 95.5\,\mathrm{GeV}$ at $\sqrt{s} = 250\,\mathrm{GeV}$ (red) and $200~\mathrm{GeV}$ (blue) within the recoil mass window $93.5\,\mathrm{GeV} \leq M_{\rm recoil} \leq 97.5~\mathrm{GeV}$. (b) Discovery region for a 95\,GeV scalar in the $\kappa_Z$–Br($S \to b\bar{b}$) plane at $\sqrt{s} = 250\,\mathrm{GeV}$ and $\mathcal{L} = 5\,\mathrm{ab}^{-1}$, with the LEP-preferred region shown in red, the SM-like Higgs branching ratio as a blue line, and the gray area is excluded by the di-photon signal strength.
}
\end{figure*}

A selection cut of $93.5\,\mathrm{GeV} \leq M_{\rm recoil} \leq 97.5\,\mathrm{GeV}$ is applied to both signal and background events, before and after DNN classification. The signal significance is then calculated for both cases. In~\autoref{fig:significance}, we show the variation of significance with integrated luminosity, ranging from $50\,\mathrm{fb}^{-1}$ to $500\,\mathrm{fb}^{-1}$. The significance is notably improved after applying the DNN. For $\mathcal{L} = 500\,\mathrm{fb}^{-1}$,~\autoref{tab:cutflow} presents a significance of $11\sigma$ ($8\sigma$) for $\sqrt{s} = 250\,\mathrm{GeV}$ ($200\,\mathrm{GeV}$) before DNN classification, which increases to $19\sigma$ ($14\sigma$) after DNN application.

We further present the discovery prospects for a scalar boson of mass 95\,GeV at a future $e^+ e^-$ collider. In a model-independent approach, we take into account its branching ratio to a bottom-quark pair and its coupling to the $Z$ boson, parametrized by the factor $\kappa_Z$. In~\autoref{fig:discovery}, we show the discovery regions in the $\kappa_Z$ {\it{vs}} Br($S \rightarrow b \bar{b}$) plane for $\sqrt{s} = 250\,\mathrm{GeV}$ at integrated luminosities of $\mathcal{L} = 0.1\,\mathrm{ab}^{-1}$ and $5\,\mathrm{ab}^{-1}$, together with the allowed region from LEP measurements and the exclusion regions from $h \rightarrow \gamma \gamma$ decays under the assumption that $S$ is SM-like. It is evident from the figure that a discovery is possible within the 95\% confidence-level region preferred by the LEP excess.

\section{Conclusion}
We study the prospects of a new scalar particle ($S$) with mass 95\,GeV at a future $e^+ e^-$ collider with center-of-mass energies of 200\,GeV and 250\,GeV. We demonstrate the effectiveness of DNN classification, which increases the signal significance within the window $93.5\,\mathrm{GeV} \leq M_{\rm recoil} \leq 97.5~\mathrm{GeV}$ by 70\% (75\%) for $\sqrt{s} = 250~\mathrm{GeV}$ (200 GeV). The CEPC, operating at a center-of-mass energy of 240\,GeV–250\,GeV, is expected to provide an integrated luminosity of $5\,\mathrm{ab}^{-1}$ over a period of 10 years, which would require approximately 100 days {of data taking} to achieve a $5\sigma$ discovery. The use of DNN classification can reduce this time, enabling the discovery in just 34 days. In this study, we consider only the $Z \rightarrow \mu^+ \mu^-$ decay channel; the results can be further improved by including additional $Z$ decay channels.

\bibliographystyle{unsrt}

\end{document}